\documentclass[useAMS,referee]{biom}
\let\bibhang\relax
 
\usepackage{natbib}
\usepackage{slashbox}
\usepackage{amsmath}
\usepackage{multirow}
\usepackage{rotating}
\def\bSig\mathbf{\Sigma}

\RequirePackage[colorlinks,citecolor=blue,urlcolor=blue]{hyperref}

\title[Using Survival Information in Truncation by Death Problems]{Using Survival Information in Truncation by Death Problems Without the Monotonicity Assumption}

\author
{Fan Yang\emailx{fan.3.yang@ucdenver.edu} \\
Department of Biostatistics and Informatics, University of Colorado Denver, Aurora, CO 80045, U.S.A.
\and
Peng Ding\emailx{pengdingpku@berkeley.edu} \\
Department of Statistics, University of California, Berkeley, CA 94720, U.S.A. 
}

\begin{document}

%  This will produce the submission and review information that appears
%  right after the reference section.  Of course, it will be unknown when
%  you submit your paper, so you can either leave this out or put in 
%  sample dates (these will have no effect on the fate of your paper in the
%  review process!)

%  These options will count the number of pages and provide volume
%  and date information in the upper left hand corner of the top of the 
%  first page as in published papers.  The \pagerange command will only
%  work if you place the command \label{firstpage} near the beginning
%  of the document and \label{lastpage} at the end of the document, as we
%  have done in this template.

%  Again, putting a volume number and date is for your own amusement and
%  has no bearing on what actually happens to your paper!  

%\volume{}
%\pubyear{}
%\artmonth{}
%
%%  The \doi command is where the DOI for your paper would be placed should it
%%  be published.  Again, if you make one up and stick it here, it means 
%%  nothing!
%
%\doi{10.1111/j.1541-0420.2005.00454.x}

%  This label and the label ``lastpage'' are used by the \pagerange
%  command above to give the page range for the article.  You may have 
%  to process the document twice to get this to match up with what you 
%  expect.  When using the referee option, this will not count the pages
%  with tables and figures.  

\label{firstpage}

%  put the summary for your paper here

\begin{abstract}
In some randomized clinical trials, patients may die before the measurements of their outcomes. Even though randomization generates comparable treatment and control groups, the remaining survivors often differ significantly in background variables that are prognostic to the outcomes. This is called the truncation by death problem. Under the potential outcomes framework, the only well-defined causal effect on the outcome is within the subgroup of patients who would always survive under both treatment and control. Because the definition of the subgroup depends on the potential values of the survival status that could not be observed jointly, without making strong parametric assumptions, we cannot identify the causal effect of interest and consequently can only obtain bounds of it. Unfortunately, however, many bounds are too wide to be useful. We propose to use detailed survival information before and after the measurements of the outcomes to sharpen the bounds of the subgroup causal effect. Because survival times contain useful information about the final outcome, carefully utilizing them could improve statistical inference without imposing strong parametric assumptions. Moreover, we propose to use a copula model to relax the commonly-invoked but often doubtful monotonicity assumption that the treatment extends the survival time for all patients. 
\end{abstract}

%  Please place your key words in alphabetical order, separated
%  by semicolons, with the first letter of the first word capitalized,
%  and a period at the end of the list.
%

\begin{keywords}
Bounds; Causal inference; Principal stratification; Survivor average causal effect.
\end{keywords}

%  As usual, the \maketitle command creates the title and author/affiliations
%  display 

\maketitle

%  If you are using the referee option, a new page, numbered page 1, will
%  start after the summary and keywords.  The page numbers thus count the
%  number of pages of your manuscript in the preferred submission style.
%  Remember, ``Normally, regular papers exceeding 25 pages and Reader Reaction 
%  papers exceeding 12 pages in (the preferred style) will be returned to 
%  the authors without review. The page limit includes acknowledgements, 
%  references, and appendices, but not tables and figures. The page count does 
%  not include the title page and abstract. A maximum of six (6) tables or 
%  figures combined is often required.''

%  You may now place the substance of your manuscript here.  Please use
%  the \section, \subsection, etc commands as described in the user guide.
%  Please use \label and \ref commands to cross-reference sections, equations,
%  tables, figures, etc.
%
%  Please DO NOT attempt to reformat the style of equation numbering!
%  For that matter, please do not attempt to redefine anything!

\section{Introduction}
\label{s:intro}
In randomized clinical studies, evaluations of the effectiveness of alternative treatments on a non-mortality outcome such as the health related quality of life (HRQOL) outcome are often complicated by truncation by death. The motivation of our study comes from a prostate cancer research \citep{petrylak2004docetaxel}, a Southwest Oncology Group (SWOG) clinical trial, where an interest is to assess the effect on a HRQOL outcome measured at six months after treatment among advanced refractory prostate cancer patients being treated with Docetaxel and Estramustine (DE) versus Mitoxantrone and Prednisone (MP). But some patients died within six months after treatment, and therefore their HRQOL outcomes were not measured and were not even well-defined. As it is well known, to estimate the treatment effect on a HRQOL outcome that is truncated by death, a direct comparison between survivors in two treatment arms could be biased. This is because death serves as a mechanism of informative censoring given its strong correlation with the HRQOL. Intuitively, those patients who died would usually have had worse HRQOL outcomes than those who survived had they somehow been kept alive \citep{cox1992quality}. To formulate a well-defined treatment effect on outcomes truncated by death, we adopt the principal stratification framework \citep{frangakis2002principal}. Based on whether each patient would survive to the HRQOL outcome measurement under treatment and whether the patient would survive to the HRQOL outcome measurement under control, subjects are classified into four principal strata as will be discussed in Section \ref{assumptions}. The stratum of patients who would survive to the outcome measurement under both treatment and control are called always survivors. Following \citet {frangakis2002principal} and \citet{rubin2006}, we focus on the treatment effect among the always survivors, also called the survivor average causal effect (SACE), because they have well-defined HRQOL outcomes under both treatment arms.

Unfortunately, the SACE is not pointly identified without strong and untestable assumptions such as by imposing a full parametric model \citep{zhang2009likelihood, frumento2012evaluating} or by utilizing a substitution variable for the latent survival type \citep{ding2011identifiability, jiang2016principal, ding2017principal, wang2017identification}. Although plausible for the settings considered by the authors, those assumptions can be too strong to be applied to many general cases where the parametric constraint is questionable and where a valid substitution variable is not available. Under weaker assumptions on the degree of selection bias, one can instead achieve partial identification on the SACE. \citet{zhang2003estimation} derived large sample bounds on the SACE under ranked average score assumptions, which are the shortest bounds possible without further assumptions. \citet{long2013sharpening} sharpened the bounds by using covariates. \citet{yang2016using} extended the ranked average score assumptions to further utilize survival information at a time point after the measurement of the HRQOL outcome. 
%\citet{hudgens2003analysis}, \citet{nolen2011randomization} and \citet{wang2017causal} developed tests on the SACE. Some other authors considered inference on the SACE through sensitivity analysis \citep{gilbert2003sensitivity, hayden2005estimator, egleston2007causal, ding2017principal}.

In the previous literature on bounding the SACE, we are aware of two limitations. First, to shorten the bounds on the SACE without imposing a parametric model, the monotonicity assumption on the survival status is often invoked with a few exceptions \citep{zhang2003estimation, ding2011identifiability, wang2017identification}. The monotonicity assumption states that the treatment does not cause death compared to the control, meaning that if a subject would die before the measurement of the HRQOL outcome therefore being truncated under treatment, the subject would also die before the measurement of the HRQOL outcome had the subject received the control. This assumption cannot be directly validated, and can be suspicious in many studies. Taking our motivating prostate cancer study SWOG as an example, although DE results in longer survival on average \citep{petrylak2004docetaxel}, from a clinical point of view, the monotonicity assumption that no patients could benefit from MP in survival may not be true since both DE and MP are active treatments. Previous research \citep{ding2011identifiability} suggested that the monotonicity assumption was likely to be violated in the SWOG study. Second, survival tends to be positively correlated with HRQOL and is informative as a proxy of health condition, but the previous literature on bounding the SACE utilizes only limited information of survival. \citet{zhang2003estimation}, by imposing the ranked average score assumptions, utilized the survival information right before the measurement on the HRQOL outcome, and showed that this information was helpful to sharpen the bounds of the SACE. The authors derived the bounds with and without imposing the monotonicity assumption. \citet{yang2016using} showed that using a post measurement time point survival information in addition to the survival information right before the measurement could further improve the bounds on the SACE. However, in practice, more detailed survival information is often available in studies where patients were followed up for multiple times, such as the SWOG clinical trial where patients were followed up at three months, six months and twelve months, respectively. This more detailed survival information can provide additional information on the health condition of the patients, and therefore, help further improve the inference on the SACE. In addition, the method the authors developed relied on the monotonicity assumption, which limited the applicability of the method.

In this paper, under the principal stratification framework, we propose a set of ranked average score assumptions to incorporate detailed survival information to sharpen the inference on the SACE in the context of randomized trials, and meanwhile, remove the untestable and often violated monotonicity assumption on survival. Because only one potential survival status is observed for each patient, the stratum membership is not observable in general. With detailed survival information, the four principal strata can be further divided into many finer ones given the potential values of the survival time, thereby introducing additional complications to inference. To address the issue of latent stratum membership, assumptions on the joint distribution of potential survivals are necessary. In our approach, we model the joint distribution of the two potential survival times under treatment and control through a copula \citep{nelsen2006introduction}. Copulas provide a flexible way for modeling dependencies and have been used to model the joint distributions of potential outcomes in various contexts \citep{bartolucci2011modeling, ma2011causal, conlon2017surrogacy}. Using a copula model, we avoid the monotonicity assumption, and characterize the association between the two potential survival times by a single copula parameter. For each fixed value of the copula parameter, deriving the bounds on the SACE under the proposed ranked average score assumptions defines a linear programming problem. The final bounds on the SACE can then be obtained by varying the value of the copula parameter in a plausible range. We apply our proposed method to the SWOG study. By utilizing detailed survival information, we are able to substantially narrow the bounds on the SACE for the effect of DE versus MP.

\section{Notation and Assumptions}
\subsection{Notation: potential and observed outcomes}
We consider two-arm randomized experiments and adopt the potential outcomes framework to define causal effects. We let $D_i$ be the binary treatment for the $i$-th subject $(i=1,...,n)$; we call level $1$ ``the treatment'' and level $0$ ``the control''. Let $\bmath{D}$ denote the vector of treatment assignment indicators for all subjects. We use $S_i(\bmath{d})$ to represent the discretized potential survival time of subject $i$ from the initiation of the treatment that would be observed under treatment assignment $\bmath{d}$ which could be measured, for instance, by month, by year, etc. Assuming that there are $K$ follow-ups at time points $s_1, s_2, ..., s_K,$ respectively from the initiation of the treatment, then the values that $S_i(\bmath{d})$ can take are $s_0, s_1, ..., s_K$ where $s_0=0$. Take the SWOG study as an example, subjects were followed up three times, and therefore $K=3$. $S_i(\bmath{d})$ will take a value $0~(s_0=0)$ if under the treatment assignment $\bmath{d}$ patient $i$ would die before the first follow-up time three months, $3~ (s_1=3)$ if patient $i$ would die between three months and the second follow up time six months, $6~(s_2=6)$ if patient $i$ would die between six months and the last follow up time twelve months, and $12~(s_3=12)$ if patient $i$ would still be alive at twelve months. We use $Y_i(\bmath{d})$ to denote the binary potential HRQOL outcome of subject $i$ that would be observed under treatment assignment $\bmath{d}$. We consider HRQOL outcome measured at a fixed time point, the $T$th follow-up ($T<K$), and therefore for subjects who would die before time point $s_T$, i.e., those with $S_i(\bmath{d})<s_T$, their potential HRQOL outcomes $Y_i(\bmath{d})$'s are not defined. Throughout this paper, we let level $1$ of the HRQOL outcome be worse than level $0$. We use $S_i$ and $Y_i$ to denote respectively subject $i$'s observed survival time and observed HRQOL outcome.

\subsection{Assumptions and the parameter of interest}
\label{assumptions}

%We invoke the following assumptions.
%\assumption{Stable unit treatment value assumption (SUTVA).}\label{ass1} Let $\bmath{d}$ and $\bmath{d}'$ be any two possible treatment assignments. If $d_i = d'_i$, then $S_i(\bmath{d}) = S_i(\bmath{d}')$, and $Y_i(\bmath{d}) = Y_i(\bmath{d}')$. Therefore, $S_i(\bmath{d})$ and $Y_i(\bmath{d})$ could be written as $S_i(d_i)$ and $Y_i(d_i)$, respectively.
%
%SUTVA means that there is only a single version of each treatment level and that there is no interference between subjects so that one subject's potential outcomes are unaffected by other subjects' treatment assignments. This assumption is plausible in many studies, such as the SWOG trial where a patient's health outcomes are unlikely to be affected by other patients' treatments received.  
%
%Under SUTVA, 

Under the stable unit treatment value assumption (SUTVA), there is only a single version of each treatment level and there is no interference between subjects. Therefore, we can write $S_i(\bmath{d})$ and $Y_i(\bmath{d})$ as $S_i(d_i)$ and $Y_i(d_i)$, respectively. Moreover, the subjects can be classified into four latent groups based on the joint values of potential survival status at the HRQOL outcome measurement time $s_T$ under treatment and under control. Let $U_i$ denote subject $i$'s latent group, which is defined as: $U_i = $ always survivor if $S_i(1)\geq s_T$ and $S_i(0)\geq s_T$, meaning that the subject would survive at least to the time point of measurement under both treatment and control; $U_i = $ protected if $S_i(1)\geq s_T$ and $S_i(0)<s_T$, meaning that the subject would survive at least to the time point of measurement only under treatment; $U_i = $ harmed if $S_i(1)<s_T$ and $S_i(0)\geq s_T$, meaning that the subject would survive at least to the time point of measurement only under control; and $U_i = $ never survivor if $S_i(1)<s_T$ and $S_i(0)<s_T$, meaning that the subject would die before the time point of measurement under both treatment and control. Among those four groups, the always survivors constitute the only group for which both $Y_i(1)$ and $Y_i(0)$ are well defined at time $s_T$. Thus the treatment effect on the HRQOL outcome is only well defined for always survivors \citep{frangakis2002principal, rubin2006}, that is, the survivor average causal effect:
\begin{equation}
 \textsc{SACE}=\mathrm{E}\{Y_i(1)-Y_i(0) \mid S_i(1)\geq s_T, S_i(0)\geq s_T\}.
\end{equation}

\assumption \label{ass2} The treatment $D_i$ is independent of the potential outcomes $S_i (1)$, $S_i (0)$, $Y_i (1)$ and $Y_i(0)$.

In randomized studies such as SWOG, the ignorable treatment assignment assumption is guaranteed by design. 

The following two assumptions compare two groups of subjects: $ G_1 = \{i\mid S_i(0) = s_{t_0}, S_i(1) = s_{t_1}\}$ and $G_2  = \{i\mid S_i(0) = s_{t'_0}, S_i(1) = s_{t'_1}\}$, where $t_0, t_1, t'_0, t'_1\in \{0, 1, 2, ..., T,..., K\}$.
\assumption \label{ass3} (i) When both groups $G_1$ and $G_2$ have well defined $Y_i(1)$ (i.e., $t_1\geq T$ and $t'_1\geq T$), if $s_{t_1}\geq s_{t'_1} $ and $s_{t_0}\geq s_{t'_0}$, then, $P\{Y_i(1)=1\mid S_i(0) = s_{t_0}, S_i(1) = s_{t_1}\}\leq P\{Y_i(1)=1\mid S_i(0) = s_{t'_0}, S_i(1) = s_{t'_1}\}$;\\
(ii) When both groups $G_1$ and $G_2$ have well defined $Y_i(0)$ (i.e., $t_0\geq T$ and $t'_0\geq T$), if $s_{t_0}\geq s_{t'_0}$ and $s_{t_1}\geq s_{t'_1}$, then, $P\{Y_i(0)=1\mid S_i(0) = s_{t_0}, S_i(1) = s_{t_1}\}\leq P\{Y_i(0)=1\mid S_i(0) = s_{t'_0}, S_i(1) = s_{t'_1}\}$.

Assumption \ref{ass3} compares the HRQOL outcomes between two groups of subjects where one group's survival times under treatment and control are both longer than or equal to those of the other group. The probability of a worse HRQOL outcome for the group with longer survival times is not higher than that for the other group, recalling that level $1$ of the HRQOL outcome is worse than level $0$.

\assumption\label{ass4} (i) When both groups $G_1$ and $G_2$ have well defined $Y_i(1)$ (i.e., $t_1\geq T$ and $t'_1\geq T$), if $s_{t_1}\geq s_{t'_1}, s_{t_0} < s_{t'_0}$, but $s_{t_1}-s_{t'_1}\geq s_{t'_0}-s_{t_0}$, then, $P\{Y_i(1)=1\mid S_i(0) = s_{t_0}, S_i(1) = s_{t_1}\}\leq P\{Y_i(1)=1\mid S_i(0) = s_{t'_0}, S_i(1) = s_{t'_1}\}$; \\(ii) When both groups $G_1$ and $G_2$ have well defined $Y_i(0)$ (i.e., $t_0\geq T$ and $t'_0\geq T$), if $s_{t_0}\geq s_{t'_0}, s_{t_1} < s_{t'_1}$, but $s_{t_0}-s_{t'_0}\geq s_{t'_1}-s_{t_1}$, then $P\{Y_i(0)=1\mid S_i(0) = s_{t_0}, S_i(1) = s_{t_1}\}\leq P\{Y_i(0)=1\mid S_i(0) = s_{t'_0}, S_i(1) = s_{t'_1}\}$.

Assumption \ref{ass4}(i) compares the HRQOL outcome under treatment between two groups of subjects where one group has longer survival under treatment but shorter survival under control than the other group. If one group's additional length of survival under treatment compared to the other group is no less than their reduced length of survival under control, Assumption \ref{ass4}(i) says that the probability of the worse HRQOL outcome under treatment for the group with longer survival under treatment is not higher than that for the other group. Assumption \ref{ass4}(ii) is the analogous assumption on the HRQOL outcome under control.

Assumptions \ref{ass3} and \ref{ass4} are our generalized ranked average score assumptions which utilize the survival information on multiple time points. They are plausibly satisfied in many HRQOL studies because survival is often positively related to the HRQOL. Therefore, it is reasonable to assume that the potential survival time is positively associated with a better potential HRQOL outcome, and is more predictive to the potential HRQOL outcome under the same treatment condition than under a different treatment condition. In particular, Assumption \ref{ass3} says that the subjects with longer survivals under both treatment and control tend to be healthier on average, and therefore, are less likely to develop the bad HRQOL outcome. Assumption \ref{ass4}(i) says that for subjects' health condition under treatment, survival under treatment is a better predictor of the HRQOL than survival under control. Therefore, even if one group of subjects live shorter than the other group under control, as long as they live much longer under treatment, they are healthier under treatment and are less likely to develop bad HRQOL outcome under treatment. Similarly, for subjects' health condition under control, Assumption \ref{ass4}(ii) says that it is survival under control is a better predictor.

\textit{Remark} $1$. Our generalized ranked average score assumptions on the HRQOL become intuitive if we consider the following generalized linear models:
\begin{eqnarray*}
 P\{Y_i(1)=1\mid S_i(0)=s_{t_0}, S_i(1)=s_{t_1}\} &=& g^{-1}( \beta-\beta_0 s_{t_0}-\beta_1 s_{t_1} ) , \\ 
 P\{Y_i(0)=1\mid S_i(0)=s_{t_0}, S_i(1)=s_{t_1}\} &=& g^{-1}( \gamma-\gamma_0 s_{t_0}-\gamma_1 s_{t_1} ) .
\end{eqnarray*}

If $g(x) = \log(x)$, then $e^{-\beta_d}$ and $e^{-\gamma_d}$ are the conditional relative risks of $S_i(d)$ on the treatment and control potential HRQOL for $d=0, 1.$ The models above imply that
\begin{eqnarray}
\log  \frac{P\{   Y_i(1)=1\mid S_i(0) = s_{t_0}, S_i(1) = s_{t_1}  \} }{P\{   Y_i(1)=1\mid S_i(0) = s_{t_0'}, S_i(1) = s_{t_1'}  \} }
&=& -\beta_0 (s_{t_0} - s_{t_0'}) - \beta_1 (s_{t_1} - s_{t_1'}) , \label{eq1} \\
\log  \frac{P\{   Y_i(0)=1\mid S_i(0) = s_{t_0}, S_i(1) = s_{t_1}  \} }{P\{   Y_i(0)=1\mid S_i(0) = s_{t_0'}, S_i(1) = s_{t_1'}  \} }
&=& -\gamma_0 (s_{t_0} - s_{t_0'}) - \gamma_1 (s_{t_1} - s_{t_1'}) . \label{eq2}
\end{eqnarray}
If $\beta_0, \beta_1, \gamma_1, \gamma_0 \geq  0$, then Assumption \ref{ass3} holds. If $\beta_1 \geq \beta_0$, i.e., the conditional relative risk of $S_i(1)$ on $Y_i(1)$ is larger than $S_i(0)$, then Assumption \ref{ass4}(i) holds. If $\gamma_0 \geq  \gamma_1$, i.e., the conditional relative risk of $S_i(0)$ on $Y_i(0)$ is larger than $S_i(1)$, then Assumption \ref{ass4}(ii) holds.  

If $g(x)=x$, then $-\beta_d$ and $-\gamma_d$ are the partial regression coefficients of $S_i(d)$ on the treatment and control potential HRQOL for $d=0, 1.$ Then the right-hand sides of \eqref{eq1} and \eqref{eq2} are the corresponding comparisons of the conditional probabilities on the risk difference scale.  
If $\beta_0, \beta_1, \gamma_1, \gamma_0 \geq  0$, then Assumption \ref{ass3} holds. If $\beta_1 \geq \beta_0$, i.e., the partial regression coefficient of $S_i(1)$ on $Y_i(1)$ is larger than $S_i(0)$, then Assumption \ref{ass4}(i) holds. If $\gamma_0 \geq  \gamma_1$, i.e., the partial regression coefficient of $S_i(0)$ on $Y_i(0)$ is larger than $S_i(1)$, then Assumption \ref{ass4}(ii) holds.  

Note that the models in this remark are used to aid interpretations, and our method does not need to invoke them.

\citet{zhang2003estimation}, hereafter ZR, proposed ranked average score assumptions to utilize the survival information on a single time point, i.e., the survival status when the HRQOL outcome was measured. Their assumptions say that when assigned to treatment, the risk of the worse HRQOL outcome for always survivors is not higher than that for the protected; and when assigned to control, the risk of the worse HRQOL outcome for always survivors is not higher than that for the harmed. Mathematically, ZR assumed,
\begin{eqnarray}
P\{Y_i(1)=1\mid S_i(1)\geq s_T, S_i(0)\geq s_T\} &\leq& P\{Y_i(1)=1\mid S_i(1)\geq s_T, S_i(0)< s_T\},  \label{zr1}\\
P\{Y_i(0)=1\mid S_i(1)\geq s_T, S_i(0)\geq s_T\} &\leq& P\{Y_i(0)=1\mid S_i(1)< s_T, S_i(0)\geq s_T\}. \label{zr2}
\end{eqnarray}
Different from our assumptions, ZR used only the survival information at the time point of the measurement of the HRQOL outcome. However, detailed survival information contains additional information on the health condition of the subjects, which can help further sharpen the inference on the SACE. Moreover, detailed survival information creates strata finer than the four principal strata, and the comparisons among those finer strata result in more plausible assumptions than the coarse comparisons among the four principal strata. For example, consider the SWOG study of the effect of DE (treatment) versus MP (control) on the HRQOL at six months described in the introduction, where patients were followed up at three months, six months and twelve months. Let us compare the following two groups' HRQOL outcomes. The first group consists of patients who would die shortly after six months no matter being treated by DE or MP, i.e., $S_i(1)=S_i(0)=6$. The second group consists of patients who would survive more than twelve months no matter being treated by DE or MP, i.e., $S_i(1)=S_i(0)=12$. Apparently, the first group subjects' health conditions at six months are much worth than those of the second group because the first group would die shortly after six months no matter being treated by DE or MP. Therefore, it's reasonable to assume that the first group's risk of bad HRQOL outcome is not lower than that for the second group as implied by our Assumption \ref{ass3}. However, ZR did not provide a comparison between two groups like those. For another example, ZR assumed that the protected, on average, had worse HRQOL outcomes than always survivors under treatment. In contrast, our Assumption \ref{ass4} assumes that some particular subgroups of always survivors, on average, have worse HRQOL outcomes than some particular subgroups of the protected under treatment, which are more reasonable assumptions for many HRQOL studies given more survival information. Let's still consider the SWOG study and compare the following two groups' HRQOL outcomes under DE. The first group again consists of patients who would die shortly after six months no matter being treated by DE or MP, i.e., $S_i(1)=S_i(0)=6$. The second group consists of patients who would survive more than twelve months if being treated by DE and would die shortly after three months if being treated by MP, i.e., $ S_i(1)=12$, and $S_i(0)=3$. Patients respond differently to different treatments. Although the patients in the second group do not respond to MP as those patients in the first group, they respond to DE much better than the patients in the first group. Thus, it is reasonable to assume that when being treated by DE, the second group of patients, a subgroup of the protected, are likely to be less sick at six month, and therefore, their risks of bad HRQOL outcome are not higher than those for the first group of patients, a subgroup of always survivors. 

\citet{yang2016using}, hereafter YS, extended ZR's ranked average score assumptions by utilizing a post measurement time point survival information. Their assumptions could be viewed as a simple version of our generalized ranked average score assumptions with two follow-up time points and with monotonicity constraints on the survival. The authors considered a scenario where the subjects were followed up twice ($K=2$) and their HRQOL outcomes were measured at the first follow up time point ($T=1$). Besides the SUTVA and Ignorability assumptions, the monotonicity assumption was imposed to restrict the possible combinations of the values of $S_i(1)$ and $S_i(0)$. Mathematically, the monotonicity assumption states that, 
$
S_i(1)\geq S_i(0).
$
As we discussed in the introduction, this assumption is strong and often suspicious. Assumptions $5$ to $7$ in YS are exactly the same as our generalized ranked average score assumptions (i.e., Assumptions \ref{ass3} and \ref{ass4}) in this special case of $K=2,~T=1$ and equal lengths of follow-up intervals (i.e., $s_2-s_1 = s_1$) or longer second follow-up interval than the first (i.e., $s_2-s_1> s_1$), without considering subjects whose survival would be harmed by the treatment. Compared with YS, our assumptions are more conservative in the case of longer first follow-up interval than the second (i.e., $s_2-s_1<s_1$). Consider two groups of patients where the first group consists of patients who would survive to the first follow-up time point $s_1$ under both treatment and control (i.e., patients with $S_i(1)=S_i(0)=s_1$), and the second group consists of patients who would survive at least to the second follow-up time point $s_2$ under treatment however would die even before the first follow-up time point (i.e., patients with $S_i(1)=s_2, S_i(0)=0$). YS always assume that the second group's risk of bad HRQOL outcome under treatment is not higher than that for the first group; in contrast, we only make this assumption when the second group's additional length of survival under treatment compared to the first group (i.e., $s_2-s_1$) is no less than their reduced length of survival under control (i.e., $s_1$). In addition, the derived bounds in YS rely on the monotonicity assumption, which may or may not cover the true effect when the monotonicity is violated. See the numerical examples in the Supplementary Materials. 

\section{Derivation of Bounds Under a Copula Model}

In this section, we derive large sample bounds for the SACE under Assumptions \ref{ass2}--\ref{ass4} assuming that the observable joint distribution of $(D_i, S_i, Y_i)$ is known. 
\subsection{Bounds given the joint distribution of $S_i(1)$ and $S_i(0)$}
\label{bounds}
Define $q_{t_1t_0d}=P\{Y_i(d)=1\mid S_i(1)=s_{t_1}, S_i(0)=s_{t_0}\}$. Define $p_{t_1 t_0} =P\{S_i(1)=s_{t_1}, S_i(0)=s_{t_0}\}$ as the proportion of the fine stratum that consists of patients who would survive to time $s_{t_1}$ (i.e., $t_1$-th follow-up) under treatment and $s_{t_0}$ (i.e., $t_0$-th follow-up) under control. In this subsection, we assume that the $p_{t_1 t_0}$'s are known. In terms of the fine strata, the SACE is:
\begin{align}\label{obj}
\textsc{SACE} &= \mathrm{E}\{Y_i(1)-Y_i(0)\mid S_i(1)\geq s_{T}, S_i(0)\geq s_T\}\nonumber\\&= P\{Y_i(1)=1\mid S_i(1)\geq s_T, S_i(0)\geq s_T\}-P\{Y_i(0)=1\mid S_i(1)\geq s_T, S_i(0)\geq s_T\}\nonumber\\&=\frac{\sum_{t_1=T}^{K}\sum_{t_0= T}^{K}(q_{t_1 t_0 1}-q_{t_1 t_0 0})p_{t_1 t_0}}{\sum_{t_1=T}^{K}\sum_{t_0= T}^{K}p_{t_1 t_0}}.
\end{align}
Meanwhile, under Assumption \ref{ass2}, $P(Y_i=1, S_i=s_t\mid D_i=d) = P\{Y_i(d)=1, S_i(d)=s_t\}$. Therefore, the observable distribution $P(Y_i, S_i\mid D_i)$ is a mixture distribution of potential outcomes of fine strata as shown in the following identities: 
\begin{equation}\label{linear1}
P(Y_i=1, S_i=s_{t_0}\mid D_i=0) = \sum_{t_1=0}^{K}p_{t_1 t_0} q_{t_1 t_0 0}, ~\textit{for}~\forall t_0\geq T,~\textit{and}
\end{equation}
\begin{equation}\label{linear2}
P(Y_i=1, S_i=s_{t_1}\mid D_i=1) = \sum_{t_0=0}^{K}p_{t_1 t_0} q_{t_1 t_0 1},  ~\textit{for}~\forall t_1\geq T. 
\end{equation}
If we hypothetically know the proportions of each fine stratum $p_{t_1 t_0}$, then the bounds for the SACE could be obtained by solving a linear programming problem with objective function (\ref{obj}), subject to the linear equality constraints (\ref{linear1}) and (\ref{linear2}), the linear inequality constraints (\ref{linearb}) that all the probabilities $q_{t_1 t_0 d}$'s are bounded between $0$ and $1$, 
\begin{equation}
0\leq q_{t_1 t_0 d} \textit{'s} \leq 1\label{linearb},
\end{equation}
and the linear inequality constraints (\ref{linear3}) and (\ref{linear4}) imposed by Assumptions \ref{ass3} and \ref{ass4}:
\begin{eqnarray}
q_{t_1 t_0 1}\leq q_{t_1't_0'1}, &&~\textit{for all~} t_1\geq t_1'\geq T, s_{t_1}+s_{t_0}\geq s_{t_1'}+s_{t_0'}\label{linear3}, \\
q_{t_1 t_0 0}\leq q_{t_1't_0'0}, &&~\textit{for all~} t_0\geq t_0'\geq T, s_{t_1}+s_{t_0}\geq s_{t_1'}+s_{t_0'}\label{linear4}.
\end{eqnarray}

\subsection{A copula model}\label{sec::copulamodel}
Although the marginal distributions of the potential survival times, $P\{S_i(1)=s_t\}$ and $P\{S_i(0)=s_t\}$ are observable, their joint distribution is not. Therefore, the proportions of fine strata $p_{t_1 t_0}$'s are not identified without further assumptions. To capture the dependence between the two potential survival times, we propose to use the copula. We assume that the joint distribution of these intermediate outcomes follows a Plackett copula \citep{plackett1965class}, where the degree of association is measured by a single parameter. Let $F_d(\cdot)$ be the marginal distribution function for the random variable $S_i(d)$ with $F_d(s_t) = P\{S_i(d)\leq s_t\}$ $(d = 0, 1)$. The joint distribution function $F(s_{t_1}, s_{t_0})$ of $S_i(1)$ and $S_i(0)$ linked by a Plackett copula is given by $C_{\phi}(F_1(s_{t_1}), F_0(s_{t_0}))$, where $C_{1}(u,v) =  uv$ when $\phi=1$, and $C_{\phi}(u,v) =\frac{\{1+(\phi-1)(u+v)\}-[\{1+(\phi-1)(u+v)\}^2-4\phi(\phi-1)uv]^{1/2}}{2(\phi-1)}$ when $\phi>0$ and $\phi\neq 1. $ The parameter $\phi$ measures the association between $S_i(1)$ and $S_i(0)$, and the Spearman correlation coefficient $\rho$ is a monotonic function of $\phi$: $\rho=\frac{\phi+1}{\phi-1}-\frac{2\phi\log{\phi}}{(\phi-1)^2}$. Therefore, $S_i(1)$ and $S_i(0)$ are independent for $\phi=1$, negatively associated for $\phi<1$, and positively associated for $\phi>1$. Since the survival times of patients under both treatment arms are highly dependent on their underlying health status, it is reasonable to assume that patients who live longer under one treatment arm are more likely to live longer under the other, implying that $\phi\geq1$ (i.e., $\rho\geq 0$). 

For a fixed $\phi$, the $p_{t_1 t_0}$'s could be calculated based on the joint distribution function, $F(s_{t_1}, s_{t_0}) = C_{\phi}(F_1(s_{t_1}), F_0(s_{t_0}))$. Then given the values of the $p_{t_1t_0}$'s, the bounds for the SACE can be obtained by solving the linear programming problem described in Section \ref{bounds}. The final bounds for the SACE will be constructed by varying the value of $\phi$ on a suitable grid and obtaining the bounds  for each value of $\phi$. The final lower bound will then take the value of the smallest lower bound among all the lower bounds, and the final upper bound will take the value of the largest upper bound among all the upper bounds. Alternatively, we can view $\phi$ as a sensitivity parameter, and draw conclusions at different values of $\phi$. 

We give two numerical examples in the Supplementary Materials to show the improvement of our approach over ZR's and the potential bias of YS's when monotonicity does not hold. Although the intuition is overwhelming that our method will lead to narrower bounds than ZR's in many cases, it is technically  challenging to give a formal proof. Even under monotonicity, YS did not give a formal proof of improvement over ZR although in most cases the improvement is apparent.

\section{Statistical Inference Accounting for Sampling Variability}\label{inference}
The bounds derived in the previous section are large sample bounds, where we assume that the joint distribution of $(Y_i, S_i)$ under each treatment arm is known. However, in practice, we need to account for the sampling uncertainty in statistical inference. Because our bounds are results of a linear programming problem, they will be in the form of intersection (i.e., the lower/upper bound takes the maximum/minimum
a collection of functionals). The maximum and minimum operators involved in the intersection bounds generate significant complications for both estimation and inference from a frequentist perspective.
%: the sample analogue estimates of the bounds have finite sample biases \citep{manski2000monotone, manski2009more}, the asymptotic distributions of these estimators are difficult to establish, and the standard resampling technique may yield inconsistent confidence intervals \citep{andrews2000inconsistency, andrews2009validity, romano1989bootstrap, romano2008inference, romano2010inference}. In addition, given the limited sample size, the estimated distribution of $(D_i,Y_i, S_i)$ based on the resampled data is not guaranteed to be always compatible with the proposed assumptions even when the assumptions hold. Other 
Most methods \citep[e.g.][]{chernozhukov2013intersection} focusing on asymptotic properties may not have desirable finite sample properties as investigated by \citet{yang2016using}. A recent method proposed by \citet{jiang2017using} requires explicit forms of the bounds. To avoid these difficulties in frequentists' inference, we adopt a Bayesian approach to conduct inference by deriving the exact credible intervals for the bounds. The Bayesian approaches are being increasingly adopted in partially identified models \citep{gustafson2009interval, scharfstein2011analysis, mealli2013using, gustafson2015bayesian}.
% \citep{gustafson2009interval, liao2010bayesian, scharfstein2011analysis, moon2012bayesian, mealli2013using}. 

Suppose that the subjects are followed up to time point $K$. Let us define parameter vectors $\boldsymbol{\pi}_d = (\pi_{0d},\pi_{1d},...,\pi_{Kd})$ for $d=0, 1$ where $\pi_{td}=P\{S_i(d)=s_t\}$. Given the value of the copula parameter $\phi$, the joint probability for survival times under treatment and control is, $
p_{t_1t_0} = C_{\phi}(F_1(s_{t_1}), F_0(s_{t_0}))-C_{\phi}(F_1(s_{t_1-1}), F_0(s_{t_0}))-C_{\phi}(F_1(s_{t_1}), F_0(s_{t_0-1}))+C_{\phi}(F_1(s_{t_1-1}), F_0(s_{t_0-1})),
$
where $F_d(s_t) = \sum_{j=0}^{j=t} \pi_{jd}$, and $F_d(s_{-1})$ is defined as $0$ for $d=0, 1$. We define parameters $\alpha_{td} = P\{Y(d)=1\mid S(d)=s_t\}$ for $d=0, 1$ and $t=T, T+1,...,K$. Given parameters $\boldsymbol{\pi}_d$'s and $\alpha_{td}$'s, the joint probability of survival and HRQOL outcomes under each treatment arm is, $P\{Y=1, S=s_t\mid D=d\}=P\{Y(d)=1, S(d)=s_t\}=\pi_{td}\alpha_{td}$, for $d=0, 1$ and $t=T, T+1,...,K$. We further define compatible region to be the region of parameters constrained by Assumptions \ref{ass2}--\ref{ass4}, i.e., 
\begin{align}
\textit{Compatible Region} = &\{\boldsymbol{\pi}_0, \boldsymbol{\pi}_1, \alpha_{td}\textit{'s}~ (d=0, 1, t=T, T+1, ..., K) \mid \textit{there are feasible solutions}\nonumber\\ &q_{t_1 t_0 1}\textit{'s}~ (T\leq t_1 \leq K, 0 \leq t_0\leq K) ~\textit{and} ~q_{t_1 t_0 0}\textit{'s} ~(0\leq t_1 \leq K, T \leq t_0\leq K) \nonumber\\& \textit{to the linear constraints}~ (\ref{linear1}) - (\ref{linear4}) \textit{defined in Section 3.1}\}
\end{align}

Prior to observing any data, we assume that the $\boldsymbol{\pi}_d$'s are independent and follow truncated Dirichlet distributions with all parameters being $1$, i.e., $f(\boldsymbol{\pi}_d)$ $\propto I(\textit{Compatible Region})\cdot \mathrm{Dirichlet}(1,...,1)$, where $I(\cdot)$ is the indicator function taking on the value $1$ if the statement is true and $0$ otherwise. Similarly, we assume that the $\alpha_{td}$'s are independent and follow truncated Beta distributions with all parameters being $1$, i.e., $f(\alpha_{td})$ $\propto I(\textit{Compatible Region})\cdot \mathrm{Beta}(1,1)$. The posterior distributions of the $\boldsymbol{\pi}_d$'s and $\alpha_{td}$'s can be derived analytically according to the priors and the likelihood. As an illustration, details on the posterior distributions calculation are provided for the SWOG study in Section \ref{sec::application}. 

To find the corresponding joint posteriors of the bounds given the value of the copula parameter $\phi$, we simulate from the posterior distributions of the $\boldsymbol{\pi}_d$'s and $\alpha_{td}$'s, and perform the linearly constrained optimizations to obtain the lower and upper bounds on the SACE for each simulation. Credible intervals for the bounds are not unique. A $100(1-\alpha)\%$ credible interval for the bounds on the SACE would be any interval where there is a $100(1-\alpha)\%$ posterior probability that the bounds (i.e., both the lower and upper bounds) fall within. Among all the $100(1-\alpha)\%$ credible intervals for the bounds on the SACE given the value of $\phi$, we choose the one with the shortest length by a numerical search.

\section{Application to the Southwest Oncology Group Study}\label{sec::application}

The data previously analyzed by \citet{ding2011identifiability} contain 487 androgen-independent prostate cancer patients enrolled in the Southwest Oncology Group (SWOG) trial between 1999 and 2003, who were randomized to receive either Docetaxel and Estramustine (DE) or Mitoxantrone and Prednisone (MP). We view the patients who received DE as the treatment group ($D_i=1$), and the patients who received MP as the control group ($D_i=0$). Let $Y_i=1$ if there was a reduction in patient $i$'s HRQOL six months after treatment compared to his HRQOL at baseline (i.e., a worse HRQOL outcome), and $Y_i=0$ otherwise. If the patient died before six months after treatment, the HRQOL level would not be measured and $Y_i$ will be undefined. Note that we dichotomize the continuous HRQOL measurement to be a binary outcome indicating its reduction compared to its baseline. Although it might cause loss of information, this dichotomized outcome is a meaningful measure of the efficacy of the treatment. Moreover, for statistical inference of this binary outcome, we do not need to impose any modeling assumptions in contrast to the original continuous outcome.

As described in Section 2, $S_i$ takes values $s_0=0, ~s_1=3,~s_2= 6$ or $s_3=12$ for death before three months, between three months and six months, between six months and twelve months, or after twelve months, respectively. The corresponding $T$ for the Southwest Oncology Group study equals $2$, and the corresponding $K$ equals $3$. Among the patients, there are 135 of them have missing measurements for their HRQOL although survived beyond six months. In our analysis, we assume that $Y_i$ is missing at random given survival $S_i$ and the treatment assignment $D_i$, and therefore, we can ignore the missing data model in the Bayesian analysis.
We set the priors to be independent and non-informative as described in Section \ref{inference}, $f(\boldsymbol{\pi}_1,\boldsymbol{\pi}_0, \alpha_{21},\alpha_{31},\alpha_{20},\alpha_{31})\propto I(\textit{Compatible Region}).$
Let $N_{ytd} = \sum_{i=1}^nI(Y_i=y, S_i=s_t, D_i=d)$ for $y=0, 1$, $t=2, 3$ and $d=0, 1$, $U_{td} = \sum_{i=1}^{n}I(Y_i ~\textit{undefined}, S_i=s_t, D_i=d)$ for $t=0, 1$ and $d=0, 1$, and let $M_{td}=\sum_{i=1}^nI(Y_i~\textit{missing}, S_i=s_t, D_i=d)$ for $t=2, 3$ and $d=0, 1$. Under the missing at random assumption, the posterior density is
\begin{align}
&~~~~f(\boldsymbol{\pi}_1,\boldsymbol{\pi}_0, \alpha_{21},\alpha_{31},\alpha_{20},\alpha_{30}\mid \boldsymbol{Y},\boldsymbol{S},\boldsymbol{D})\nonumber\\&\propto I(\textit{Compatible Region})\cdot \pi_{01}^{U_{01}}\cdot\pi_{11}^{U_{11}}\cdot\pi_{21}^{M_{21}+N_{121}+N_{021}}
\cdot \pi_{31}^{M_{31}+N_{131}+N_{031}}\nonumber\\&~~~\cdot \pi_{00}^{U_{00}}\cdot\pi_{10}^{U_{10}}\cdot\pi_{20}^{M_{20}+N_{120}+N_{020}}
\cdot \pi_{30}^{M_{30}+N_{130}+N_{030}}\nonumber\\&~~~\cdot \alpha_{21}^{N_{121}}\cdot (1-\alpha_{21})^{N_{021}}\cdot \alpha_{31}^{N_{131}}\cdot (1-\alpha_{31})^{N_{031}}\cdot\alpha_{20}^{N_{120}}\cdot (1-\alpha_{20})^{N_{020}}\cdot \alpha_{30}^{N_{130}}\cdot (1-\alpha_{30})^{N_{030}},\nonumber
\end{align}
where $\bmath{Y}$, $\bmath{S}$, and $\bmath{D}$ are the observed vectors of the HRQOL outcomes, survival times and treatment assignment indicators for all subjects.
Therefore, the posterior distributions of $\boldsymbol{\pi}_1$ and $\boldsymbol{\pi}_0$ are Dirichlet with parameters $(U_{01}+1,U_{11}+1, M_{21}+N_{121}+N_{021}+1, M_{31}+N_{131}+N_{031}+1)$ and $(U_{00}+1,U_{10}+1, M_{20}+N_{120}+N_{020}+1, M_{30}+N_{130}+N_{030}+1)$, respectively, and the posterior distributions of $\alpha_{21},\alpha_{31},\alpha_{20},\alpha_{30}$ are Beta with parameters $(N_{121}+1, N_{021}+1)$, $(N_{131}+1, N_{031}+1)$, $(N_{120}+1, N_{020}+1)$, $(N_{130}+1, N_{030}+1)$, respectively, all truncated within the compatible region. Truncation could be done by simulating from the un-truncated distributions and rejecting the simulations without feasible solutions to the linear constrained optimization problem. The details of the optimization problem we solve for the SWOG study are presented in the Supplementary Materials. 

\begin{sidewaystable}
\caption{Results for the SWOG study comparing our general ranked average score assumptions with ZR's ranked average score assumptions.}
    \centering
\begin{tabular}{c c | c c | c c }
\hline
$\rho$&$\log(\phi)$&Estimated Bounds & Relative Length& $95\%$ Credible Interval&  Relative Length \\[0.5ex] 
\hline
0.0000 & 0.000 & $[-0.110, 0.005]$& 0.764 & $[-0.226, 0.128]$ & 0.889\\
0.1000 & 0.301 & $[-0.105, -0.002]$& 0.684 & $[-0.222, 0.116]$ & 0.850\\
0.2000 & 0.607 & $[-0.099, -0.008]$& 0.602 & $[-0.216, 0.106]$ & 0.811\\
0.3000 & 0.926 & $[ -0.092,-0.015]$& 0.515 & $[-0.204, 0.103]$ & 0.770\\
0.4000 & 1.264 & $[-0.085, -0.020]$& 0.428 & $[-0.194, 0.095]$ & 0.727\\
0.5000 & 1.632 & $[-0.077, -0.025]$& 0.346 & $[-0.187, 0.086]$ & 0.685\\
0.6000 & 2.049 & $[-0.070, -0.029]$& 0.274 & $[-0.180, 0.080]$ & 0.654\\
0.7000 & 2.545 & $[-0.065, -0.031]$& 0.224 & $[-0.174, 0.078]$ & 0.633\\
0.8000 & 3.189 & $[-0.062, -0.033]$& 0.191 & $[-0.167, 0.079]$ & 0.620 \\
0.9000 & 4.191 & $[-0.059, -0.037]$& 0.148 & $[-0.168, 0.073]$ & 0.606\\
0.9900 & 7.110 & $[-0.053, -0.042]$& 0.072 & $[-0.162, 0.069]$ & 0.582\\
0.9990 & 9.773 & $[-0.052, -0.043]$& 0.056 & $[-0.161, 0.069]$ & 0.578\\
0.9999 &12.331 &$[-0.052, -0.043]$& 0.056 & $[-0.161, 0.069]$ & 0.578\\
\hline
ZR& & $[-0.130, 0.020]$&1.0000 &$[-0.241, 0.157]$ & 1.0000\\

\hline %inserts single line
\end{tabular}\label{result}
\end{sidewaystable}

Table \ref{result} compares the estimated bounds and the $95\%$ credible intervals under the following two sets of assumptions: (i) Assumptions \ref{ass2}--\ref{ass4}, and (ii) Assumptions \ref{ass2} with ZR's ranked average score assumptions (\ref{zr1}) and (\ref{zr2}). The point estimates of the lower and upper bounds are reported based on their posterior medians, and the posterior distributions of the bounds under (ii) are obtained similarly with non-informative Beta priors on parameters $P\{S_i(d)\geq 6\}$'s and $P\{Y_i(d)=1\mid S_i(d)\geq 6\}$'s for $d=0, 1$. According to the results under the set of assumptions (i), with a moderately small correlation between the two potential survival times under DE and MP, $(e.g., \rho\geq 0.2)$, among the patients with androgen-independent prostate cancer who would survive to at least six months under both DE and MP, DE would help reduce the risk of bad HRQOL by 0.8 percent to 9.9 percent. Given the facts that the two potential survival times and two potential HRQOL outcomes are highly dependent on subjects' underlying health status, and that the correlation between the potential HRQOL outcome and potential survival time under DE is 0.22, $\rho\geq 0.2$ is probably a plausible and conservative assumption on the correlation between the two potential survival times under DE and MP. These bounds are about $40\%$ shorter and are also more informative than the bounds obtained by utilizing only the survival information at the time point of measurement, which estimates that the effect of DE is somewhere between reducing the risk of bad HRQOL outcome by 13.0 percent and increasing the risk by 2 percent compared to MP. Unfortunately, due to the small sample size and missing data, all the $95\%$ credible intervals cover $0$, indicating that there is not enough evidence to conclude that DE improves the HRQOL outcome among androgen-independent prostate cancer patients who would be able to survive to at least six months under both treatments. 

To evaluate the sensitivity of the proposed approach to the choice of the type of copula, we also conducted the analysis using the Gaussian copula. The results are very close. For instance, with a correlation $\rho\geq 0.2$ between the two potential survival times under DE and MP, under the Gaussian copula, the estimated bounds are $[-0.098, -0.008]$ (compared to $[-0.099, -0.008]$ under the Plackett copula) with a $95\%$ credible interval $[-0.216, 0.107]$ (compared to $[-0.216, 0.106]$ under the Plackett copula). To save space, the description of the Gaussian copula and the detailed results are presented in the Supplementary Materials.    

\section{Discussion}

Our approach can also be applied to stratified randomized trials where strata are created based on prognostic factors and randomization is conducted within each stratum. In these settings, we can weaken our assumptions by requiring them to hold in each stratum. Each stratum specific SACE can be bounded, and the overall bounds on the SACE will then be obtained as a weighted average of stratum specific SACE's with weights proportional to the sizes of strata measured by the numbers of always survivors \citep{freiman2014large}. 

We focused on binary outcomes which allow for obtaining nonparametric bounds of the SACE. If the original outcome $Y^*$ is continuous and dichotomization will lose information, we can consider applying our method to estimate SACE on $Y = I(Y^*>y)$ at each point of $y$, or imposing additional modeling assumptions on $Y^*.$ We leave this to future research. 

In our analysis of the SWOG data, we focus on the patients' HRQOL outcome measured at six months after initiation of the treatment. However, patients' HRQOL outcomes were measured repeatedly in this study at each follow-up time point, namely, three months, six months and twelve months. Using the current approach, one would conduct separate analyses on the SACE's for the HRQOL outcomes measured at different time points, which is not ideal to study the change in the SACE over time. How to incorporate the information from repeated measurements of the HRQOL outcomes requires further research.

Subject matter knowledge is necessary to judge the plausibility of our assumptions because they could not be validated by the observable data. We expect that our approach could be widely applied to HRQOL outcomes because patients surviving longer tend to be healthier, and patients' health conditions under one arm tend to be better predicted by their survival lengths under the same arm than that under a different arm. However, in studies where the relationship between the time to truncation and the outcome of interest is not clear, our assumptions could be suspicious. Consider the causal effect of a job-training program on participants' wages that are potentially truncated by unemployment. For some subjects, longer time to get employed may indicate their being less competitive in the job market, and therefore, is associated with lower wages. However, for some subjects that are very competitive and can afford longer unemployment, they may decline some job opportunities with low wages, and therefore, longer time to get employed is associated with higher wages. In such a case, our assumptions may not apply, and alternative assumptions should be invoked to sharpen the inference.

\section{Supplementary Materials}
Numerical examples referenced in Section \ref{sec::copulamodel}, details on the optimization problem and sensitivity analysis for the SWOG study referenced in Section \ref{sec::application}, original data and source code can be found at the {\it Biometrics} website on Wiley Online Library.

\backmatter
\section*{Acknowledgments}

We thank the Editor (Professor Stijn Vansteelandt), Associate Editor and a reviewer for constructive comments. We thank Dylan Small for helpful discussions and suggestions. Fan Yang's research is partially supported by National Institutes of Health grant R03 CA208387-01A1. Peng Ding's research was partially supported by the Institute of Education Sciences grant R305D150040.

%  We greatly prefer that you incorporate the references for your
%  article into the body of the article as we have done here 
%  (you can use natbib or not as you choose) than use BiBTeX,
%  so that your article is self-contained in one file.
%  If you do use BiBTeX, please use the .bst file that comes with 
%  the distribution.  In this case, replace the thebibliography
%  environment below by 
%
\bibliographystyle{biom} 
\bibliography{mybibiloc}

\begin{thebibliography}{}

\bibitem[\protect\citeauthoryear{Bartolucci and Grilli}{Bartolucci and
  Grilli}{2011}]{bartolucci2011modeling}
Bartolucci, F. and Grilli, L. (2011).
\newblock Modeling partial compliance through copulas in a principal
  stratification framework.
\newblock {\em Journal of the American Statistical Association} {\bf 106,}
  469--479.

\bibitem[\protect\citeauthoryear{Chernozhukov, Lee, and Rosen}{Chernozhukov
  et~al.}{2013}]{chernozhukov2013intersection}
Chernozhukov, V., Lee, S., and Rosen, A.~M. (2013).
\newblock Intersection bounds: estimation and inference.
\newblock {\em Econometrica} {\bf 81,} 667--737.

\bibitem[\protect\citeauthoryear{Conlon, Taylor, and Elliott}{Conlon
  et~al.}{2017}]{conlon2017surrogacy}
Conlon, A., Taylor, J., and Elliott, M. (2017).
\newblock Surrogacy assessment using principal stratification and a gaussian
  copula model.
\newblock {\em Statistical methods in medical research} {\bf 26,} 88--107.

\bibitem[\protect\citeauthoryear{Cox, Fitzpatrick, Fletcher, Gore,
  Spiegelhalter, and Jones}{Cox et~al.}{1992}]{cox1992quality}
Cox, D.~R., Fitzpatrick, R., Fletcher, A., Gore, S., Spiegelhalter, D., and
  Jones, D. (1992).
\newblock Quality-of-life assessment: can we keep it simple?
\newblock {\em Journal of the Royal Statistical Society: Series A (Statistics
  in Society)} {\bf 155,} 353--393.

\bibitem[\protect\citeauthoryear{Ding, Geng, Yan, and Zhou}{Ding
  et~al.}{2011}]{ding2011identifiability}
Ding, P., Geng, Z., Yan, W., and Zhou, X.-H. (2011).
\newblock Identifiability and estimation of causal effects by principal
  stratification with outcomes truncated by death.
\newblock {\em Journal of the American Statistical Association} {\bf 106,}
  1578--1591.

\bibitem[\protect\citeauthoryear{Ding and Lu}{Ding and
  Lu}{2017}]{ding2017principal}
Ding, P. and Lu, J. (2017).
\newblock Principal stratification analysis using principal scores.
\newblock {\em Journal of the Royal Statistical Society: Series B (Statistical
  Methodology)} {\bf 79,} 757--777.

\bibitem[\protect\citeauthoryear{Frangakis and Rubin}{Frangakis and
  Rubin}{2002}]{frangakis2002principal}
Frangakis, C.~E. and Rubin, D.~B. (2002).
\newblock Principal stratification in causal inference.
\newblock {\em Biometrics} {\bf 58,} 21--29.

\bibitem[\protect\citeauthoryear{Freiman and Small}{Freiman and
  Small}{2014}]{freiman2014large}
Freiman, M.~H. and Small, D.~S. (2014).
\newblock Large sample bounds on the survivor average causal effect in the
  presence of a binary covariate with conditionally ignorable treatment
  assignment.
\newblock {\em The International Journal of Biostatistics} {\bf 10,} 143--163.

\bibitem[\protect\citeauthoryear{Frumento, Mealli, Pacini, and Rubin}{Frumento
  et~al.}{2012}]{frumento2012evaluating}
Frumento, P., Mealli, F., Pacini, B., and Rubin, D.~B. (2012).
\newblock Evaluating the effect of training on wages in the presence of
  noncompliance, nonemployment, and missing outcome data.
\newblock {\em Journal of the American Statistical Association} {\bf 107,}
  450--466.

\bibitem[\protect\citeauthoryear{Gustafson}{Gustafson}{2015}]{gustafson2015bayesian}
Gustafson, P. (2015).
\newblock {\em Bayesian Inference for Partially Identified Models: Exploring
  the Limits of Limited Data}, volume 140.
\newblock CRC Press.

\bibitem[\protect\citeauthoryear{Gustafson and Greenland}{Gustafson and
  Greenland}{2009}]{gustafson2009interval}
Gustafson, P. and Greenland, S. (2009).
\newblock Interval estimation for messy observational data.
\newblock {\em Statistical Science} {\bf 24,} 328--342.

\bibitem[\protect\citeauthoryear{Jiang and Ding}{Jiang and
  Ding}{ress}]{jiang2017using}
Jiang, Z. and Ding, P. (In press).
\newblock Using missing types to improve partial identification with missing
  binary outcomes.
\newblock {\em Annals of Applied Statistics} .

\bibitem[\protect\citeauthoryear{Jiang, Ding, and Geng}{Jiang
  et~al.}{2016}]{jiang2016principal}
Jiang, Z., Ding, P., and Geng, Z. (2016).
\newblock Principal causal effect identification and surrogate end point
  evaluation by multiple trials.
\newblock {\em Journal of the Royal Statistical Society: Series B (Statistical
  Methodology)} {\bf 78,} 829--848.

\bibitem[\protect\citeauthoryear{Long and Hudgens}{Long and
  Hudgens}{2013}]{long2013sharpening}
Long, D.~M. and Hudgens, M.~G. (2013).
\newblock Sharpening bounds on principal effects with covariates.
\newblock {\em Biometrics} {\bf 69,} 812--819.

\bibitem[\protect\citeauthoryear{Ma, Roy, and Marcus}{Ma
  et~al.}{2011}]{ma2011causal}
Ma, Y., Roy, J., and Marcus, B. (2011).
\newblock Causal models for randomized trials with two active treatments and
  continuous compliance.
\newblock {\em Statistics in medicine} {\bf 30,} 2349--2362.

\bibitem[\protect\citeauthoryear{Mealli and Pacini}{Mealli and
  Pacini}{2013}]{mealli2013using}
Mealli, F. and Pacini, B. (2013).
\newblock Using secondary outcomes to sharpen inference in randomized
  experiments with noncompliance.
\newblock {\em Journal of the American Statistical Association} {\bf 108,}
  1120--1131.

\bibitem[\protect\citeauthoryear{Nelsen}{Nelsen}{2006}]{nelsen2006introduction}
Nelsen, R.~B. (2006).
\newblock {\em An Introduction to Copulas (2nd ed.)}.
\newblock New York: Springer.

\bibitem[\protect\citeauthoryear{Petrylak, Tangen, Hussain, Lara~Jr, Jones,
  Taplin, et~al\mbox{.}}{Petrylak et~al.}{2004}]{petrylak2004docetaxel}
Petrylak, D.~P., Tangen, C.~M., Hussain, M.~H., Lara~Jr, P.~N., Jones, J.~A.,
  Taplin, M.~E., et~al. (2004).
\newblock Docetaxel and estramustine compared with mitoxantrone and prednisone
  for advanced refractory prostate cancer.
\newblock {\em New England Journal of Medicine} {\bf 351,} 1513--1520.

\bibitem[\protect\citeauthoryear{Plackett}{Plackett}{1965}]{plackett1965class}
Plackett, R.~L. (1965).
\newblock A class of bivariate distributions.
\newblock {\em Journal of the American Statistical Association} {\bf 60,}
  516--522.

\bibitem[\protect\citeauthoryear{Rubin}{Rubin}{2006}]{rubin2006}
Rubin, D.~B. (2006).
\newblock Causal inference through potential outcomes and principal
  stratification: application to studies with ``censoring" due to death.
\newblock {\em Statistical Science} {\bf 21,} 299--309.

\bibitem[\protect\citeauthoryear{Scharfstein, Onicescu, Goodman, and
  Whitaker}{Scharfstein et~al.}{2011}]{scharfstein2011analysis}
Scharfstein, D., Onicescu, G., Goodman, S., and Whitaker, R. (2011).
\newblock Analysis of subgroup effects in randomized trials when subgroup
  membership is missing: application to the second multicenter automatic
  defibrillator intervention trial.
\newblock {\em Journal of the Royal Statistical Society: Series C (Applied
  Statistics)} {\bf 60,} 607--617.

\bibitem[\protect\citeauthoryear{Wang, Zhou, and Richardson}{Wang
  et~al.}{2017}]{wang2017identification}
Wang, L., Zhou, X.-H., and Richardson, T.~S. (2017).
\newblock Identification and estimation of causal effects with outcomes
  truncated by death.
\newblock {\em Biometrika} {\bf 104,} 597--612.

\bibitem[\protect\citeauthoryear{Yang and Small}{Yang and
  Small}{2016}]{yang2016using}
Yang, F. and Small, D.~S. (2016).
\newblock Using post-outcome measurement information in censoring-by-death
  problems.
\newblock {\em Journal of the Royal Statistical Society: Series B (Statistical
  Methodology)} {\bf 78,} 299--318.

\bibitem[\protect\citeauthoryear{Zhang and Rubin}{Zhang and
  Rubin}{2003}]{zhang2003estimation}
Zhang, J.~L. and Rubin, D.~B. (2003).
\newblock Estimation of causal effects via principal stratification when some
  outcomes are truncated by death.
\newblock {\em Journal of Educational and Behavioral Statistics} {\bf 28,}
  353--368.

\bibitem[\protect\citeauthoryear{Zhang, Rubin, and Mealli}{Zhang
  et~al.}{2009}]{zhang2009likelihood}
Zhang, J.~L., Rubin, D.~B., and Mealli, F. (2009).
\newblock Likelihood-based analysis of causal effects of job-training programs
  using principal stratification.
\newblock {\em Journal of the American Statistical Association} {\bf 104,}
  166--176.

\end{thebibliography}

\label{lastpage}

\newpage

\begin{center}
\Huge Supplementary Material
\end{center}

\section{Numerical Examples}
\label{example}

\subsection{Numerical example showing improvements}

Consider a hypothetical study where subjects are followed up four times ($K=4$) at time points $1$, $2$, $3$, and $4$ with equal follow-up intervals and the HRQOL outcome is measured at the second follow-up ($T=2$). We set the marginal distributions $\pi_{td} = P\{S(d) = s_t\}$ for $t=0$ to $4$ and $d=0, 1$ as follows:
\begin{align}
&\pi_{01} = 0.15, ~~~~\pi_{11} = 0.25, ~~~~\pi_{21}=0.20,~~~~\pi_{31}=0.25, ~~~~\pi_{41} = 0.15,\nonumber\\
&\pi_{00} = 0.15, ~~~~\pi_{10} = 0.15, ~~~~\pi_{20}=0.30,~~~~\pi_{30}=0.15, ~~~~\pi_{40} = 0.25.\nonumber
\end{align}
Given the marginal distributions, their joint distribution is specified by a Plackett copula with $\phi=7.76$, corresponding to a Spearman correlation $\rho=0.6$. The underlying risks of developing the bad HRQOL outcome under treatment and under control for each fine stratum are described in Table \ref{risk}. 

\begin{table}[hb]
\caption{Risks of the bad HRQOL outcome for each fine stratum with $S_i(1)=t_1$ and $S_i(0)=t_0$ under treatment ($q_{t_1t_0 1}$) and under control ($q_{t_1t_0 0}$) in the numerical example in Section 1.1. The risks are presented in the form of $q_{t_1t_0 1} (q_{t_1t_0 0})$, with ``$-$'' indicating that the corresponding risk is not defined.}
\begin{tabular}{  l | c c c c c}
 \backslashbox{$t_1$}{$t_0$} & $0$ & $1$ & $2$ & $3$ & $4$\\
\hline
$0$ & $- ~(-)$ &$- ~(-)$  & $-~(0.60)$ & $-~(0.55)$ &$-~(0.45)$ \\
$1$ & $- ~(-)$ &$- ~(-)$ &$-~(0.55)$  & $-~(0.50)$ & $-~(0.40)$\\
$2$ & $0.80~(-)$ &$0.75~(-)$  & $0.75~(0.50)$ &$0.70~(0.45)$  & $0.65~(0.35)$\\
$3$ & $0.75~(-)$ &$0.70~(-)$ &$0.70~(0.50)$  &$0.65~(0.45)$  & $0.60~(0.35)$\\
$4$ & $0.65~(-)$ &$0.60~(-)$ &$0.60~(0.45)$  &  $0.55~(0.40)$& $0.50~(0.30)$\\
\end{tabular}\label{risk}
\end{table}
In the above setup, SACE $= 0.210$, meaning that the treatment increases the risk of the bad HRQOL outcome by $0.210$ among patients who will survive at least to time point $2$ under both treatment arms. 

Suppose that we have an infinite sample. Under Assumption 1, we would observe the values of $\pi_{td}$'s since $\pi_{td} = P(S_i=s_t\mid D_i=d)$, and would observe the joint distribution of $Y_i$ and $S_i$ in each treatment arm,
\begin{align}
P(Y_i=1, S_i=2\mid D_i=1) =0.146, ~P(Y_i=1, S_i=3\mid D_i=1) =0.163,\nonumber\\ P(Y_i=1, S_i=4\mid D_i=1) =0.079,~
P(Y_i=1, S_i=2\mid D_i=0) =0.157, \nonumber\\P(Y_i=1, S_i=3\mid D_i=0) =0.068, ~P(Y_i=1, S_i=4\mid D_i=0) =0.084.\nonumber
\end{align}

\begin{table}[ht]
\caption{Bounds on the SACE for the numerical example in Section 1.1 under two different sets of assumptions: Assumptions 1--3, and Assumptions 1 with ZR's ranked average score assumptions in (4) and (5).}
    \centering
\begin{tabular}{c c c c}
\hline
$\rho$&$\log(\phi)$&Bounds & Relative Length\\[0.5ex] 
\hline
$0.000$ & 0.000 & $[0.118, 0.299]$&0.228\\
$0.100$ & 0.301 &$[0.134, 0.284]$& 0.188 \\
0.200 & 0.607 &$[0.146, 0.271]$& 0.158\\
0.300 & 0.926 &$[0.156, 0.261]$& 0.132 \\
0.400 & 1.264 & $[0.165, 0.256]$& 0.115 \\
0.500 & 1.632 & $[0.173, 0.152]$& 0.099 \\
0.600 & 2.049 & $[0.182, 0.248]$& 0.084 \\
0.700 & 2.545 & $[0.191, 0.246]$& 0.069\\
0.800 & 3.189 & $[0.200, 0.243]$& 0.055 \\
0.900 & 4.191 & $[0.210, 0.242]$& 0.041  \\
0.990 & 7.110 & $[0.218, 0.241]$& 0.029\\
0.999 & 9.773 &$[0.219, 0.241]$& 0.028 \\
 0.9999 &12.331 &$[0.219, 0.241]$& 0.028\\
\hline
\multicolumn{2}{c}{ZR} & $[-0.147, 0.647]$& 1.000\\ 
\hline %inserts single line
\end{tabular}\label{numerical}
\end{table}

Table \ref{numerical} presents the bounds for the SACE under Assumptions 1--3 as a function of the Spearman correlation coefficient $\rho$. The bounds get tighter as $\rho$ increases. This reflects the fact that with larger  $\rho$, there is less uncertainty in each subject' stratum membership when one of the two potential survival status is observed. With no prior information on $\rho$ and by conservatively assuming that $\rho\geq 0$, we obtain the bounds for the SACE as $[0.118, 0.299]$, showing that the treatment increases the risk of the bad HRQOL outcome.

In contrast, if we utilize only the survival information at the time point of the HRQOL outcome measurement, under Assumptions 1 with ZR's ranked average score assumptions (4) and (5), the bounds on the SACE are $[-0.147, 0.647]$. Based on these bounds, we would not know whether or not the treatment increases the risk of the bad HRQOL outcome even though the true value of the SACE is positive. Table 2 also presents the relative lengths of the bounds compared to the ZR's. Under Assumptions 1--4, the width of the bounds is reduced by $77.2\%$ even with a conservative specification $\rho \geq 0$, showing that the detailed survival information helps narrow the bounds on the SACE.

The underlying distribution of this hypothetical example describes a scenario where the commonly assumed monotonicity assumption in the previous literature is violated. Under the above specification with $\rho=0.6$, the proportions of always survivors, protected, harmed, and never survivors are $0.519, ~0.081, ~0.181$ and $0.219$, respectively. When the monotonicity assumption does not hold, YS's bounds may or may not include the true effect. In this example, applying YS's bounds by utilizing time point $4$'s (a post measurement time point) survival information, the bounds on the SACE are $[0.188,~0.298]$ which cover the true effect.

\subsection{Additional numerical example showing the consequence with incorrectly assumed monotonicity assumption}

We show an example in which YS's method produces bounds that exclude the true effect, and therefore lead to biased result. Let us still consider a hypothetical study where subjects are followed up four times ($K=4$) at time points $1$, $2$, $3$, and $4$ with equal follow-up intervals and the HRQOL outcome is measured at the second follow-up ($T=2$). We set the marginal distributions $\pi_{td} = P\{S(d) = s_t\}$ for $t=0$ to $4$ and $d=0,~1$ as follows:
\begin{align}
&\pi_{01} = 0.30, ~~~~\pi_{11} = 0.40, ~~~~\pi_{21}=0.15,~~~~\pi_{31}=0.10, ~~~~\pi_{41} = 0.05,\nonumber\\
&\pi_{00} = 0.40, ~~~~\pi_{10} = 0.30, ~~~~\pi_{20}=0.15,~~~~\pi_{30}=0.10, ~~~~\pi_{40} = 0.05.\nonumber
\end{align}
As before, given the marginal distributions, we specify their joint distribution by a Plackett copula with $\phi=7.76$, corresponding to a Spearman correlation $\rho=0.6$. Under this setup, the proportions of always survivors, protected, harmed, and never survivors are $0.182, ~0.118, ~0.118$ and $0.582$, respectively. The underlying risks of developing the bad HRQOL outcome under treatment and under control for each fine stratum are described in Table \ref{risk_sup}.

\begin{table}
\caption{Risks of the bad HRQOL outcome for each fine stratum with $S_i(1)=t_1$ and $S_i(0)=t_0$ under treatment ($q_{t_1t_0 1}$) and under control ($q_{t_1t_0 0}$) in the numerical example in Section 1.2. The risks are presented in the form of $q_{t_1t_0 1} (q_{t_1t_0 0})$, with ``$-$'' indicating that the corresponding risk is not defined.}
\begin{tabular}{  l | c c c c c}
 \backslashbox{$t_1$}{$t_0$} & $0$ & $1$ & $2$ & $3$ & $4$\\
\hline
$0$ & $- ~(-)$ &$- ~(-)$  & $-~(0.70)$ & $-~(0.55)$ &$-~(0.40)$ \\
$1$ & $- ~(-)$ &$- ~(-)$ &$-~(0.70)$  & $-~(0.55)$ & $-~(0.40)$\\
$2$ & $0.75~(-)$ &$0.75~(-)$  & $0.75~(0.55)$ &$0.70~(0.40)$  & $0.65~(0.25)$\\
$3$ & $0.70~(-)$ &$0.70~(-)$ &$0.70~(0.50)$  &$0.65~(0.35)$  & $0.60~(0.20)$\\
$4$ & $0.60~(-)$ &$0.60~(-)$ &$0.60~(0.45)$  &  $0.55~(0.20)$& $0.50~(0.15)$\\
\end{tabular}\label{risk_sup}
\end{table} 

In the above setup, SACE $= 0.267$, meaning that the treatment increases the risk of the bad HRQOL outcome by $0.267$ among patients who will survive at least to time point $2$ under both treatment arms. 

Suppose that we have an infinite sample. Under Assumption $1$ in the main paper, we would observe the values of $\pi_{td}$'s since $\pi_{td} = P(S_i=s_t\mid D_i=d)$, and would observe the joint distribution of $Y_i$ and $S_i$ in each treatment arm,
\begin{align}
P(Y_i=1, S_i=2\mid D_i=1) =0.110, ~P(Y_i=1, S_i=3\mid D_i=1) =0.067,\nonumber\\ P(Y_i=1, S_i=4\mid D_i=1) =0.028,~
P(Y_i=1, S_i=2\mid D_i=0) =0.091, \nonumber\\P(Y_i=1, S_i=3\mid D_i=0) =0.042, ~P(Y_i=1, S_i=4\mid D_i=0) =0.012.\nonumber
\end{align}

\begin{table}
\caption{Bounds on the SACE for the numerical example in Section 1.2 under various sets of assumptions.}
\centering
\begin{tabular}{ccccc}
\hline
SACE& \multicolumn{3}{c}{Without Monotonicity} & Monotonicity\\
\cline{2-4}
&\multicolumn{2}{c}{Assumptions 1--3}&Assumptions 1 with& \\
&& &ZR's ranked average score assumptions &\\
\hline
0.267&$\rho=0.0$& $[0.010, ~0.477]$& $[-0.485, ~0.685]$ & 0.200\\
& $\rho=0.2$& $[0.049, ~0.460]$& & \\
& $\rho=0.6$& $[0.155, ~0.351]$& & \\
& $\rho=0.9$& $[0.178, ~0.277]$& & \\
\hline
\end{tabular}\label{bounds}
\end{table}

Table \ref{bounds} presents the bounds on the SACE under various sets of assumptions. With a conservative assumption that $\rho\geq 0$, the bounds on the SACE are $[0.010,~ 0.477]$ under our generalized ranked average score assumptions, whereas ZR's ranked average score assumptions provide a much wider bounds $[-0.485, ~0.685]$ that are not informative. Both methods do not invoke the monotonicity assumption, and the bounds obtained correctly cover the true effect. In contrast, if we incorrectly impose the monotonicity assumption and apply YS's method by utilizing time point $4$'s survival information, we would conclude that the SACE is 0.200 which is $25\%$ smaller than the true effect $0.267$. The monotonicity assumption results in degenerate bounds in this example. This is because when there are no harmed subjects, the proportion of the protected is equal to the difference between the proportions of subjects survived to the measurement of the HRQOL outcome under treatment and under control, which is $0$ in this example. Then there is no uncertainty in each subject's principal strata's membership: everyone survived to the second follow-up time point is an always survivor, and everyone died before the second follow-up time point is a never survivor. 

\section{The details of the optimization problem we solve for the SWOG study}
Given the values of $\boldsymbol{\pi}_1,~\boldsymbol{\pi}_0,~\alpha_{21}, ~\alpha_{31}, ~\alpha_{20}, ~\alpha_{30}$ drawn from their posterior distribution described in the Section 5 of the main paper and the value of the copula parameter $\phi$, we first obtain the values of the joint probabilities of the survival times under treatment and under control $p_{t_1 t_0}$'s where $p_{t_1 t_0} = P(S_i(1)=s_{t_1}, S_i(0)=s_{t_0})$. In the SWOG study, recall that $K=3$, $T=2$, and $s_0=0, ~s_1=3,~s_2=6,~s_3=12$. Then, viewing $q_{t_1 t_0 d}$'s as unknown variables, we solve the following linear programming problem:

\begin{equation}
\min/\max~~~~~~ \frac{\sum_{t_1=2}^{3}\sum_{t_0= 2}^{3}(q_{t_1 t_0 1}-q_{t_1 t_0 0})p_{t_1 t_0}}{\sum_{t_1=2}^{3}\sum_{t_0= 2}^{3}p_{t_1 t_0}}\nonumber
\end{equation}
$~~Subject~~to:$\\
\begin{equation}
\sum_{t_1=0}^{3}p_{t_1 t_0} q_{t_1 t_0 0} = \pi_{t_0 0}\alpha_{t_0 0}, ~\textit{for}~ t_0 = 2, ~3\nonumber
\end{equation}
\begin{equation}
\sum_{t_0=0}^{3}p_{t_1 t_0} q_{t_1 t_0 1} = \pi_{t_1 1}\alpha_{t_1 1}, ~\textit{for}~ t_1 = 2, ~3\nonumber.
\end{equation}
\begin{equation}
0\leq q_{t_1 t_0 d}\textit{'s} \leq 1 \nonumber
\end{equation}
\begin{equation}
q_{t_1 t_0 1}\leq q_{t_1't_0'1}, ~\textit{for all~} 2\leq t_1'\leq t_1\leq 3 ~\textit{where}~ s_{t_1}+s_{t_0}\geq s_{t_1'}+s_{t_0'}\nonumber
\end{equation} 
\begin{equation}
q_{t_1 t_0 0}\leq q_{t_1't_0'0}, ~\textit{for all~} 2\leq t_0'\leq t_0\leq 3 ~\textit{where}~ s_{t_1}+s_{t_0}\geq s_{t_1'}+s_{t_0'}\nonumber
\end{equation}

\section{Results on the SWOG study using the Gaussian copula}

Let $F_d(\cdot)$ be the marginal distribution function for the random variable $S_i(d)$ with $F_d(s_t) = P\{S_i(d)\leq s_t\}$ $(d = 0, 1)$. The joint distribution function $F(s_{t_1}, s_{t_0})$ of $S_i(1)$ and $S_i(0)$ linked by the Gaussian copula is given by $C_{r}(F_1(s_{t_1}), F_0(s_{t_0}))$, where 
\begin{equation}
C_{r}(u,v) =\Phi_{r}(\Phi^{-1}(u), \Phi^{-1}(v)).\nonumber
\end{equation}
In the expression above, $\Phi^{-1}$ is the inverse of the standard normal cumulative distribution function, $\Phi_{r}$ is the joint cumulative distribution function of a bivariate normal distribution with mean vector zero and covariance matrix equal to the correlation matrix $\begin{pmatrix} 
1 & r \\
r & 1 
\end{pmatrix}$. The parameter $r$ measures the Pearson correlation between the random variables $S_i(1)$ and $S_i(0)$, and the Spearman correlation coefficient $\rho$ between $S_i(1)$ and $S_i(0)$ is a monotonic function of $r$: $\rho=\frac{6}{\pi}\cdot \mathrm{arcsin}(\frac{r}{2})$. 

We now replace the Plackett copula with the Gaussian copula in the analysis to evaluate the sensitivity of the results to the choice of the type of copula. Table \ref{sensitivity} compares the estimated bounds and the $95\%$ credible intervals under our generalized ranked average score assumptions (i.e., Assumptions $1$--$3$) by modeling the joint distribution of the potential survival times under DE and MP (i) using the Plackett copula, and (ii) using the Gaussian copula. The results are robust to the choice of the type of copula. For instance, with a correlation $\rho\geq 0.2$ between the two potential survival times under DE and MP, under the Gaussian copula, the estimated bounds are $[-0.098, -0.008]$ (compared to $[-0.099, -0.008]$ under the Plackett copula) with a $95\%$ credible interval $[-0.216, 0.107]$ (compared to $[-0.216, 0.106]$ under the Plackett copula). 

\begin{sidewaystable}
\caption{Results for the SWOG study under our generalized ranked average score assumptions comparing the Plackett copula with the Gaussian copula.}
    \centering
\begin{tabular}{c | c c | c c }
\hline
$\rho$&\multicolumn{2}{c}{Estimated Bounds} & \multicolumn{2}{c}{$95\%$ Credible Interval}\\[0.5ex] 
\cline{2-3}\cline{4-5}\\
&Plackett & Gaussian&Plackett&Gaussian\\
\hline
0.0000 &  $[-0.110, 0.005]$& [-0.110, 0.005] & $[-0.226, 0.128]$ & [-0.226, 0.128]\\
0.1000 &  $[-0.105, -0.002]$& [-0.105, -0.002] & $[-0.222, 0.116]$ & [-0.221, 0.117]\\
0.2000 &  $[-0.099, -0.008]$& [-0.098, -0.008] & $[-0.216, 0.106]$ & [-0.216, 0.107]\\
0.3000 &  $[ -0.092,-0.015]$& [-0.092, -0.014] & $[-0.204, 0.103]$ & [-0.205, 0.101]\\
0.4000 &  $[-0.085, -0.020]$& [-0.085, -0.020] & $[-0.194, 0.095]$ & [-0.196, 0.095]\\
0.5000 &  $[-0.077, -0.025]$& [-0.077, -0.025] & $[-0.187, 0.086]$ & [-0.186, 0.089]\\
0.6000 &  $[-0.070, -0.029]$& [-0.071, -0.029] & $[-0.180, 0.080]$ & [-0.177, 0.085]\\
0.7000 &  $[-0.065, -0.031]$& [-0.066, -0.031] & $[-0.174, 0.078]$ & [-0.170, 0.082]\\
0.8000 &  $[-0.062, -0.033]$& [-0.062, -0.033] & $[-0.167, 0.079]$ & [-0.168, 0.079] \\
0.9000 &  $[-0.059, -0.037]$& [-0.059, -0.036] & $[-0.168, 0.073]$ & [-0.166, 0.075]\\
0.9900 &  $[-0.053, -0.042]$& [-0.053, -0.042] & $[-0.162, 0.069]$ & [-0.163, 0.069]\\
0.9990 &  $[-0.052, -0.043]$& [-0.052, -0.043] & $[-0.161, 0.069]$ & [-0.161, 0.069]\\
0.9999 &  $[-0.052, -0.043]$& [-0.052, -0.043] & $[-0.161, 0.069]$ & [-0.161, 0.069]\\

\hline %inserts single line
\end{tabular}\label{sensitivity}
\end{sidewaystable}

\end{document}